\documentclass[preprint,showpacs,preprintnumbers,amsmath,amssymb]{revtex4}

\usepackage{graphicx}
\usepackage{dcolumn}
\usepackage{bm}

\begin{document}

\title{Shear viscosity of liquid helium 4 above the $\lambda$ point}

\author{Shun-ichiro Koh}
 
 \email{koh@kochi-u.ac.jp}
\affiliation{ Physics Division, Faculty of Education, Kochi University  \\
        Akebono-cho, 2-5-1, Kochi, 780, Japan 
}%

\date{\today}

\begin{abstract}
 In liquid helium 4, many features associated to Bose statistics have been masked by the 
strongly interacting nature of the liquid.  As an example of these 
features, we examine the shear viscosity of liquid helium 4 above the $\lambda$ point. 
  Applying the linear-response theory to Poiseuille's formula for
 the capillary flow, the  reciprocal of the shear viscosity  
  coefficient $\eta $ is considered as a transport coefficient.  
 Using the  Kramers-Kronig relation, we relate a   
 superfluid flow in a capillary with that in a rotating bucket, and 
 express $1/\eta $ in terms of the susceptibility of the system.  
 A formula for the kinematic shear viscosity $\nu $  is obtained which 
 describes the influence of Bose statistics. Using this formula, we study the gradual fall  
 of $\nu (T)$ from $3.7K$ to $T_{\lambda}$ in liquid helium 4 at 1 atm. 
 
 \end{abstract}

\maketitle

\section{\label{sec:level1}Introduction}
The shear viscosity of liquid helium 4 has been 
subjected to considerable experimental and theoretical studies.
 Well below the lambda  temperature $T_{\lambda}$, various 
excitations of a liquid are strictly suppressed except for phonons and 
rotons.  Hence, it is natural to assume that they are  
responsible for the  shear viscosity appearing in the damping of an oscillating disc, 
or the drag of a rotating cylinder at $T<T_{\lambda}$. 
By regarding these phonons and rotons as a weakly 
interacting dilute Bose gas, the kinetic theory of gases  
properly describes the shear viscosity  of liquid helium 4 at $T \ll T_{\lambda}$ \cite {kha}.

For the shear viscosity above $T_{\lambda}$, however, the 
dilute-gas picture must not be applied, because the macroscopic 
condensate, which is the basis for the above picture, has not yet developed. 
 Rather, we must deal with the influence of Bose statistics on the 
dissipation mechanism of a liquid. In an ordinary liquid flowing along x-direction, 
the shear viscosity causes the shear stress $F_{xy}$ between two adjacent 
layers at different velocities 
\begin{equation}
 F_{xy}=\eta \frac{\partial v_x}{\partial y¥}¥,
	\label{¥}
\end{equation}¥
where $\eta $ is the coefficient of shear viscosity. 
In the linear response theory, $\eta $ of a stationary flow is given by the 
following two-time correlation function of a tensor $J_{xy}(t)=-\sum_{i}(p_{i,x}p_{i,y}/m¥)$
\begin{equation}
   \eta =\frac{1}{Vk_BT¥}¥\int_{0}^{\infty¥}dt<J_{xy}(0)J_{xy}(t)>¥.
	\label{¥}
\end{equation}¥
 In principle, $\eta$ of a liquid, and therefore the effect of Bose statistics on $\eta$ is obtained by 
calculating an infinite series of the perturbation expansion of   
Eq.(2) with respect to the particle interaction $U$ \cite {kad}. 
 At first sight, however, it seems to be a hopeless attempt, because the dissipation in a liquid
 is a complicated phenomenon allowing no simple explanation \cite {han} \cite{jeo}.
 In  liquid helium 4, many features associated to Bose statistics have been masked by the 
strongly interacting nature of the liquid.

In this paper, we will go back to the original phenomenon which has been 
known from 1938.  Superfluidity was first discovered in a flow through a channel and a 
flow through a capillary \cite {kap}. 
 In an ordinary flow through a capillary  (a radius $a$ and a length $L$), 
 the velocity distribution under the pressure difference $\Delta P$ has a form such as
\begin{equation}
 v_z(r)=\frac{a^2-r^2}{4\eta¥}¥\frac{\Delta P}{L¥}¥,
	\label{¥}
\end{equation}¥
where $r$ is a radius in the cylindrical coordinates  (Poiseuille flow. Fig.1).
In the superfluid phase of liquid helium 4, even when the pressure difference vanishes ($\Delta P=0$), 
one observes a non-vanishing flow ($v_z(r) \ne 0$), hence $\eta =0$ in 
Eq.(3). Instead of Eq.(2), we will regard Poiseuille's formula Eq.(3) 
as a linear-response relation.
Without loss of generality, we may focus on a flow velocity at a single 
point, for example, 
 $\mbox{\boldmath $v$}(r=0)$ on the axis of rotational symmetry (z-axis).  
We define  a mass flow density $\mbox{\boldmath $j$}=\rho\mbox{\boldmath 
$v$}(r=0)$, and rewrite Eq.(3) as
\begin{equation}
 \mbox{\boldmath $j$}=-\sigma a^2\frac{\Delta \mbox{\boldmath $P$}}{L¥},  \qquad  \sigma =\frac{\rho}{4\eta¥}¥,
	\label{¥}
\end{equation}¥
where $\sigma$ is the conductivity of a liquid in a capillary ($\rho$ is 
a density, and $\mbox{\boldmath $P$}=P\mbox{\boldmath $e$}_z$)  \cite 
{cur}. In general, when fluid particles strongly interact  to 
each other, a fluid  swiftly responds to the shear stress, thereby having a small coefficient 
of shear viscosity $\eta $. 
In Eq.(4),  $\eta$ appears in the denominator of $\sigma$.  If we apply 
the linear-response theory to the {\it reciprocal \/} $1/\eta $, we make 
a perturbation expansion of $1/\eta$ with respect to $U$. In this case, 
{\it an increase of $U$ generally leads to an increase of $1/\eta$ and  
thereby a decrease of $\eta$ \/}.  The influence of $U$ on $\eta $ is 
naturally built into the formalism \cite {can}.

 \begin{figure}
\includegraphics [scale=0.38]{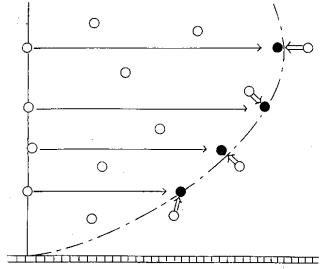}
\caption{\label{fig:epsart} A flow through a capillary.  }
\end{figure}

 The change of the starting point from Eq.(2)  to Eq.(4) opens a new 
 possibility.  Using the Kramers-Kronig relation, we will relate 
 the frictionless capillary flow with the nonclassical flow in a rotating bucket. 
This method originated in studies of electron superconductivity \cite {fer}.
Between the  above two forms of superfluidity in liquid helium 4, we 
find a parallel relationship to that between the electrical conduction and the 
 Meissner effect in superconductivity.
  This analogy permits to express $1/\eta $ in terms of the 
  susceptibility of the system, and enables to include the effect of Bose 
  statistics into the perturbation expansion.
 
As an application of this method, we examine the gradual fall of 
 $\eta (T)$ with decreasing temperature of liquid helium 4 from $3.7K$ 
 to  $T_{\lambda}$ in $1$ atm. It is often said that  this $\eta (T)$ 
 resembles $\eta (T)$ of a gas than a liquid in its magnitude and  
in its temperature dependence. But this expression is somewhat 
misleading, because the picture of a gas has no ground in this regime. 
Rather, we must consider this gradual fall of $\eta (T)$ as a property of the 
liquid under the strong influence of Bose statistics.

This paper is organized as follows.   Section 2.A considers the shear viscosity of
liquid helium 4 using Kramers-Kronig relation, and gives a formula 
for the kinematic shear viscosity $\nu (T) $.
Section.2.B discusses the physical reason for the suppression of the shear 
viscosity by Bose statistics.  Sec.3 examines the shear viscosity of liquid helium 4 
 above $T_{\lambda}$, and Sec.4 makes a comparison to experiments.
Section.5 discusses some related problems.

\section{Shear viscosity of a Bose liquid}

\subsection{Formalism}  
  For liquid helium 4, we use the following hamiltonian with the 
  repulsive interaction $U$
\begin{equation}
 H=\sum_{p}\epsilon (p)\Phi_{p}^{\dagger}\Phi_{p}
   +U\sum_{p,p'}\sum_{q}\Phi_{p-q}^{\dagger}\Phi_{p'+q}^{\dagger}\Phi_{p'}\Phi_p , 
   \qquad (U>0),¥¥¥
	\label{¥}
\end{equation}¥
where $\Phi_{p}$ denotes an annihilation operator of a spinless boson.

  Let us  generalize  Eq.(4) to the case of the oscillatory pressure as follows
\begin{equation}
 \mbox{\boldmath $j$}(\omega)=-\sigma (\omega) a^2\frac{\Delta \mbox{\boldmath $P$}(\omega)}{L¥}.
	\label{¥}
\end{equation}¥
The conductivity spectrum $\sigma (\omega)$ in Eq.(6) must 
 satisfy the following sum rule \cite {sum}
\begin{equation}
   \frac{1}{\pi ¥}¥\int_{0}^{\infty¥}\sigma (\omega)d\omega¥=f(a)¥,
	\label{¥}
\end{equation}¥
where $f(a)$ is a function determined by fluid dynamics. 
Equation (7) is a form of the oscillator-strength sum rule in terms of 
fluid conductivity through a capillary.  The Stokes equation gives an 
expression of $\sigma (\omega)$ and $f(a)$ (see Appendix.A).

\begin{figure}
\includegraphics [scale=0.5]{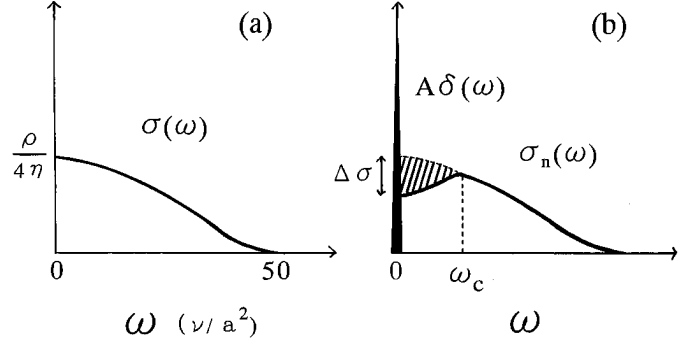}
\caption{\label{fig:epsart} The change of the conductivity spectrum $\sigma (\omega )$ from (a) at 
 $T>T_{\lambda}$ to (b) at $T<T_{\lambda}$.  At $T>T_{\lambda}$,  $\sigma (\omega )$ 
 is given by the real part of Eq.(A5).  At $T<T_{\lambda}$, the sharp peak with a half 
 width $\omega _s$ and an area $A$ appears around \protect$\omega =0$. 
 $\omega _c$  is a frequency at the  broad peak of $\sigma _n(\omega )$.  }
\end{figure}

At $T<T_{\lambda}$, a frictionless flow appears in Eq.(6). Characteristic 
to the  frictionless flow is the fact that in addition to  the normal fluid 
part $\sigma _n(\omega)$, the conductivity spectrum $\sigma (\omega)$ has 
a sharp peak at $\omega =0$ (with a very small but finite half width 
$\omega_s$). Hence, one obtains
\begin{equation}
 \mbox{\boldmath $j$}(\omega)=-\left[\sigma _n(\omega)+A\delta(\omega)\right]¥¥a^2 \frac{\Delta \mbox{\boldmath $P$}(\omega)}{L¥},
	\label{¥}
\end{equation}¥
where $\delta(\omega)$ is a simplified expression of the sharp peak at $\omega =0$, 
and $A$ is an area of this peak \cite {fer}.  Figure 2 schematically illustrates such a change of 
$\sigma (\omega)$ when the system passes $T_{\lambda}$.  
When $\sigma (\omega)$ has a form of  $\sigma _n(\omega)+A\delta(\omega)$, 
 $\eta $ in Eq.(4) is given by 
\begin{equation}
   \eta (T)=\frac{1}{4¥}¥\frac{\rho}{\sigma _n(0)+\displaystyle{\left(\frac{A}{\omega _s¥}\right)}¥¥}¥¥.
	\label{¥}
\end{equation}¥
In view of Eq.(9), the sharpness of the peak at $\omega =0$ ($\omega _s \simeq 0$) leads 
to the disappearance of the shear viscosity ($\eta (T) \simeq 0$) at $T<T_{\lambda}$.

Let us generalize  Eq.(6) so that it includes not only a mass flow  
responding in phase with $\Delta \mbox{\boldmath $P$}(\omega)$ but also 
a mass flow responding out of phase.  $\sigma (\omega)$ in Eq.(6) is 
generalized to a complex number $\sigma_1+i\sigma_2$ as follows
\begin{equation}
 \mbox{\boldmath $j$}(\omega)=-\left[\sigma_1(\omega)+i\sigma_2(\omega)\right]¥a^2 \frac{\Delta \mbox{\boldmath $P$}(\omega)}{L¥},
	\label{¥}
\end{equation}¥
where $\sigma (\omega)$ in Eq.(6) is replaced by $\sigma_1(\omega)$. 
 Instead of $\Delta \mbox{\boldmath $P$}(t)$, we will use 
 a fictitious velocity $\mbox{\boldmath $v$}(t)$ in Eq.(10).  Applying the 
 pressure gradient to a liquid is equivalent to assuming, as an external 
 field, a velocity $\mbox{\boldmath $v$}(t)$ satisfying the equation of motion  \cite {vec}
\begin{equation}
      \rho\frac{d \mbox{\boldmath $v$}(t)}{d t¥}¥= -\frac{\Delta \mbox{\boldmath $P$}(t)}{L¥} .
	\label{¥}
\end{equation}¥
When replacing $\Delta \mbox{\boldmath $P$}(\omega)/L$ with 
 $-i\rho\omega\mbox{\boldmath $v$}(\omega)$ in Eq.(10),  
the real and imaginary part of $\sigma (\omega)$ are interchanged, 
\begin{equation}
 \mbox{\boldmath $j$}(\omega)=\rho\left[-\omega\sigma_2(\omega)+i\omega\sigma_1(\omega)\right]¥¥a^2
                                                \mbox{\boldmath $v$}(\omega).
           \label{¥}
\end{equation}¥
Equation (12) is a linear-response formula, in which $\mbox{\boldmath 
$j$}(\omega)$ and $\mbox{\boldmath $v$}(\omega)$ forms a perturbation energy such as
 $\mbox{\boldmath $j$}(\omega)\cdot\mbox{\boldmath $v$}(\omega)$ \cite {vel}. The   
susceptibility $\chi(\omega)$  consists of a real part $-\omega\sigma_2(\omega)$ for the 
non-dissipative flow and an imaginary part $i\omega\sigma_1(\omega)$ for the dissipative one. 

Causality requires that one observes a mass flow only after the pressure is applied. 
We obtain the following Kramers-Kronig relation for $-\omega\sigma_2(\omega)$ and 
$\omega\sigma_1(\omega)$ in Eq.(12) 
\begin{equation}
   \sigma_1(\omega')=\frac{2}{\pi}¥\int_{0}^{\infty¥}d\omega\frac{\omega\sigma_2(\omega)}{\omega^2-\omega'^2¥}¥,
	\label{¥}
\end{equation}¥
\begin{equation}
   \omega'\sigma_2(\omega')=-\frac{2}{\pi}¥\int_{0}^{\infty¥}d\omega\frac{\omega^2\sigma_1(\omega)}{\omega^2-\omega'^2¥}¥.
	\label{¥}
\end{equation}¥
If one determines $\sigma_2(\omega)$ in the non-dissipative flow, 
 one obtains $\sigma_1(\omega)$  using Eq.(13).

  As a non-dissipative flow, we consider a flow in a rotating bucket.
 In fluid dynamics, the viscous dissipation in an incompressible fluid is 
estimated by the dissipation function 
\begin{equation}
	\Phi (\mbox{\boldmath $r$})=2\eta \left(e_{ij}-\frac{1}{3¥}e_{kk}\delta_{ij}\right)^2¥,
	\label{¥}
\end{equation}¥
($e_{ij}=(\partial v_i/\partial x_j+\partial v_j/\partial x_i)/2$ is the 
shear velocity).  Different flow patterns have  different degrees of dissipation:
  For the Poiseuille flow Eq.(3), the dissipation function 
Eq.(15) is not zero at every $r$ except for $r=0$.  
Fluid particles in a  capillary flow experience thermal 
dissipation not only at the boundary, but also inside of the flow.  
  On the other hand, in a rotating bucket, a liquid makes the rigid-body 
 rotation  owing to its viscosity.   Except at the boundary  to the wall, there is no 
 frictional force within a liquid, and the flow is therefore a 
 non-dissipative one.   (The velocity is a product of the radius and the rotational velocity $\Omega$ such as 
  $\mbox{\boldmath $v$}_d( \mbox{\boldmath $r$})\equiv \mbox{\boldmath 
 $\Omega$}\times \mbox{\boldmath $r$}$, for which Eq.(15) is zero at every $r$.)

The flow in a rotating bucket is formulated using the generalized susceptibility of 
the system  $\chi(\mbox{\boldmath $r$},\omega)$  \cite {noz}. When we 
assume $\mbox{\boldmath $v$}_d(\mbox{\boldmath $r$},t) =\mbox{\boldmath 
 $\Omega$}\times \mbox{\boldmath $r$}e^{i\omega t}$  as a 
 $\mbox{\boldmath $v$}(\omega)$ in Eq.(12), 
 $\mbox{\boldmath $j$}(\mbox{\boldmath $r$},\omega)$ and 
 $\mbox{\boldmath $v$}_d(\mbox{\boldmath $r$},\omega)$ 
form the perturbation energy like $\int dx\mbox{\boldmath 
$v$}_d\cdot \mbox{\boldmath $j$}$.  Since the flow in a rotating bucket is a non-dissipative system,   
$\mbox{\boldmath $j$}(\mbox{\boldmath $r$},\omega)$ is a dynamical response of a 
liquid to the mechanical external field $\mbox{\boldmath $v$}_d(\mbox{\boldmath $r$},\omega)$, such as 
$\mbox{\boldmath $j$}(\mbox{\boldmath $r$},\omega)= \chi(\mbox{\boldmath 
$r$},\omega)\mbox{\boldmath $v$}_d(\mbox{\boldmath $r$},\omega)$.
The influence of the wall motion propagates from 
the wall to the center along the radial direction, which is perpendicular 
to the particle motion driven by rotation. Hence, the  
flow in a rotating bucket is a transverse response described by the transverse 
susceptibility $\chi^T(q,\omega)$. In the right-hand side of Eq.(12), the 
real part $-\rho\omega\sigma_2(\omega)a^2$ is expressed as
\begin{equation}
   -\rho a^2\omega\sigma_2(\omega)=\lim_{q\to 0}\chi^T(q,\omega) ¥.
	\label{¥}
\end{equation}¥
 Using Eq.(16) in the right-hand side of Eq.(13), we obtain $\sigma_{1}(\omega)$ for the 
 capillary flow 
\begin{equation}
   \rho a^2\sigma_{1}(\omega')= -\frac{2}{\pi}¥¥
         \int_{0}^{\infty¥}d\omega\frac{\lim_{q\to 0}\chi^T(q,\omega)}{\omega^2-\omega'^2¥}¥.
	\label{¥}
\end{equation}¥

Equation(17) has the following meaning.  Quantum mechanics states that, in 
the decay from an excited state with an 
 energy level $E$ to a ground state with $E_0$, the higher excitation 
 energy $E$ causes the shorter relaxation time $\tau $. The 
 time-dependent perturbation theory says
 \begin{equation}
 	\frac{\hbar}{\tau ¥}¥ \simeq |E-E_0|.
 	\label{¥}
 \end{equation}¥
  In Eq.(17),  the left-hand side includes the relaxation time $\tau $ 
  as  $\sigma _{1} =\rho /(4\eta)=\rho /(4G\tau)$ (see Appendix.B), whereas  
 the right-hand side  includes the excitation spectrum in $\chi^T(q,\omega)$.  
 In this sense,  Eq.(17) is a many-body theoretical expression of Eq.(18).

 In the normal fluid phase, $\chi^L(q,\omega)=\chi^T(q,\omega)$ is satisfied at small $q$ and $\omega$, 
and one can replace $\chi^T(q,\omega)$ in Eq.(17) by $\chi^L(q,\omega)$
 for a small $q$ and $\omega$ \cite {def}.  Hence, the conductivity $\sigma_{1n}(\omega)$ 
 of the capillary flow in the normal fluid phase is given by 
\begin{equation}
   \rho a^2\sigma_{1n}(\omega')= -\frac{2}{\pi}¥¥
         \int_{0}^{\infty¥}d\omega\frac{\lim_{q\to 0}\chi^L(q,\omega)}{\omega^2-\omega'^2¥}¥ .
	\label{¥}
\end{equation}¥

In the superfluid phase, under the strong influence of Bose statistics, 
the condition of $\chi^L(q,\omega)=\chi^T(q,\omega)$  at $q\rightarrow 0$ is violated (see Sec.2.B).
Consequently, one cannot replace  $\chi^T(q,\omega)$ by 
$\chi^L(q,\omega)$ in Eq.(17).  In addition to $\sigma_{1n}(\omega)$, 
one must separately consider the contribution of $\chi^T(q,\omega)-\chi^L(q,\omega)$. 
Hence, one obtains  
\begin{equation}
   \sigma_1(\omega')=\sigma_{1n}(\omega')+
            \frac{2}{\rho\pi a^2¥}¥\int_{0}^{\infty¥}d\omega
            \frac{\lim_{q\to 0}[\chi^L(q,\omega)-\chi^T(q,\omega)]¥}{\omega^2-\omega'^2¥} ¥,
	\label{¥}
\end{equation}¥
in which $\lim_{q\to 0}[\chi^L(q,\omega)-\chi^T(q,\omega)]¥$ 
corresponds to a superfluid component.  In general, superfluidity is not rigid 
against external perturbations oscillating at not low frequencies: 
Above a certain $\omega _0$, $\lim_{q\to 0}[\chi^L(q,\omega)-\chi^T(q,\omega)]$ becomes zero. 
 One can see $\sigma_1(\omega)$ of such a system  using  a simple 
 example.  We assume a constant $\lim_{q\to 0}[\chi^L(q,0)-\chi^T(q,0)]$ for 
 $0\leq  \omega\leq \omega_0$, and $0$ for $\omega_0 < \omega$.  In this case, we obtain 
\begin{equation}
   \sigma_1(\omega)=\sigma_{1n}(\omega)+\frac{2}{\rho\pi a^2¥}¥\lim_{q\to 
   0}[\chi^L(q,0)-\chi^T(q,0)]D_{\omega _0}(\omega). ¥
	\label{¥}
\end{equation}¥
where 
 \begin{equation}
  D_{\omega_0}(\omega)¥=\frac{1}{2\omega¥}¥\ln|\frac{\omega _0^2}{\omega ^2¥}¥-1|.
	\label{¥}
\end{equation}¥
Figure.3 shows $D_{\omega_0}(\omega)$ for $\omega_0 =1$. For every 
$\omega_0$, $ D_{\omega_0}(\omega)$  resembles the $\delta  
(\omega)$-function. This simple example suggests that, for all probable forms of
$\lim_{q\to 0}[\chi^L(q,\omega)-\chi^T(q,\omega)]$, the sharp peak around $\omega =0$ 
is a general feature of $\sigma_1(\omega)$.  On comparing Eqs.(21) 
and (22) with Eq.(8), we obtain an expected form of $\sigma(\omega)$ for the superfluid flow,
 and $A=2\lim_{q\to 0}[\chi^L(q,0)-\chi^T(q,0)]/(\rho\pi a^2)$.  The
 half width $\omega _s$ of the peak in $\sigma_{1}(\omega)$ is determined by 
 $\omega_0$ in $\chi^L-\chi^T$, and it gives us a measure for 
 the rigidity of superfluidity.

\begin{figure}
\includegraphics [scale=0.5]{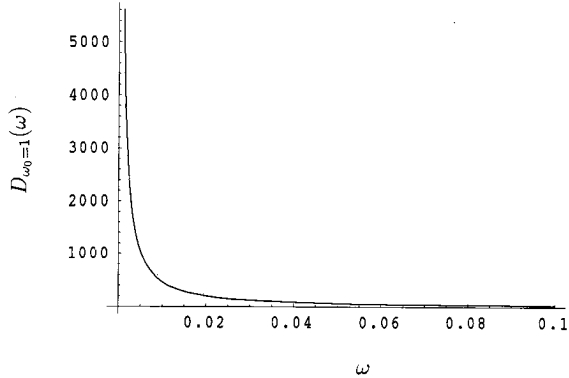}
\caption{\label{fig:epsart}  $D_{\omega_0}(\omega)$ for $\omega_0 =1$.}
\end{figure}

When the system passes $T_{\lambda}$,  $\sigma(\omega)$ changes from Fig.2(a) to 2(b) under the sum 
rule Eq.(7).  The area $A$ of the sharp peak is equal to that of the 
shaded region in Fig.2(b). At $\omega =0$, the  conductivity is enhanced 
by the sharp peak as $A/\omega _s$, 
but the original $\sigma (0)$ becomes $\sigma (0)-\Delta\sigma $ by 
the sum rule.   We approximate the shaded region in Fig.2(b) by a triangle. (A broad peak 
 of $\sigma_{n}(\omega)$ is located at $\omega_c$). We take into account 
 the sum rule as $-\Delta\sigma  \omega_c/2+A=0$, hence $\Delta\sigma =2A/\omega_c$. 
After all, the conductivity at $\omega =0$ becomes $[\sigma (0)-2A/\omega_c]+A/\omega_s$ 
at $T<T_{\lambda}$. Using this form of $\sigma (0)$ and the expression of $A$, 
one obtains $\eta =\rho/(4\sigma (0))$ as
\begin{equation}
  \eta (T)=\left(\frac{\rho}{4¥}\right)¥\frac{1}{\sigma_{1n}(0)
              +\displaystyle{\frac{2}{\rho\pi a^2¥}\left(\frac{1}{\omega_s¥}-\frac{2}{\omega_c¥}¥\right)
                                                             \lim_{q\to 0}[\chi^L(q,0)-\chi^T(q,0)]¥¥}}¥.
	\label{¥}
\end{equation}¥
Here, we define {\it the mechanical superfluid density \/} 
 $\hat {\rho _s}(T)\equiv \lim_{q\to 0}[\chi^L(q,0)-\chi^T(q,0)]$, 
  which does not always agree with the conventional thermodynamical superfluid density 
$\rho _s(T)$.  (By ``thermodynamical'', we imply the quantity that 
remains finite in the $V\rightarrow \infty $ limit.)  Using  $\omega_s¥ \ll 
\omega_c$ in Eq.(23), we obtain the following formula of {\it the kinematic shear viscosity\/} 
$\nu(T)=\eta(T)/\rho(T)$ 
\begin{equation}
  \nu (T)=¥\frac{\nu _n}{1
              +\displaystyle{\frac{8}{\pi a^2\omega_s¥}\frac{\hat {\rho _s}(T)}{\rho}}\nu _n¥¥}¥,
	\label{¥}
\end{equation}¥
where $\nu _n$ is the kinematic shear viscosity in the normal fluid 
phase, and it satisfies $\sigma_{1n}(0)=\rho/(4\eta _n)=1/(4\nu _n)$ in Eq.(23).  
(The total density  $\rho $ in Eq.(23) slightly and monotonically 
increases with decreasing temperature from $4.2K$ to $T_{\lambda}$ \cite {ker}.  
 Comparing Eq.(24) with (23), we see that $\nu (T)=\eta (T)/\rho (T)$ is a more appropriate 
quantity  than $\eta (T)$ for describing the change of the system 
around $T_{\lambda}$.)

One notes the following features in the formula (24).

(1) $\nu (T)$ of a superfluid is expressed as an infinite power series of  
$\nu _n$, and the influence of Bose statistics appears in its 
coefficients. This result does not depend on a particular model of a 
liquid, but on the general argument. (The microscopic derivation of $\nu _n$ depends on the model,  
which is a subject of the liquid theory and beyond the scope of this paper.)

(2) Because of $\omega _s\simeq 0$, which characterizes the sharp peak 
in Fig.2(b), a small change of $\hat {\rho _s}(T)$ in the 
denominator of the right-hand side is strongly enhanced to an observable change of $\nu (T)$. 

(3) The existence of $1/a^2$ in front of $\hat {\rho _s}(T)/\rho$ 
indicates that a frictionless superfluid flow appears 
only in a narrow capillary with a small radius $a$: A narrower capillary shows a clearer evidence 
of a frictionless flow.

\subsection{Physical interpretation}

There is a physical reason to expect that the shear viscosity of a Bose 
liquid falls off at low temperature. 
For the shear viscosity of a classical liquid, Maxwell obtained a simple formula 
$\eta=G\tau$ (Maxwell's relation) using a physical argument \cite{max}. ($G$ is the 
modulus of rigidity. $\tau$ is a relaxation time of the 
process in which the fluid motion relaxes the shear stress between adjacent layers in a flow. See Appendix.B). 
In the vicinity of $T_{\lambda}$, no apparent structural transformation is 
observed in liquid helium 4. Hence, $G$ may be a constant at the first 
approximation, and therefore the fall of the shear viscosity is attributed to a decrease of 
$\tau $. In view of Eq.(18), the decrease of $\tau$  suggests that the 
excitation energy $E$ increases owing to Bose statistics. The relationship between 
the excitation energy and  Bose statistics dates back to 
Feynman's argument on the scarcity of the low-energy excitation in liquid helium 
4 \cite{fey1}, in which he explained  how Bose 
statistics affects the many-body wave function in configuration space. 
To the shear viscosity, we will apply his explanation.

Consider a flow in Fig.1.  White circles represent  
an initial  distribution of fluid particles.  The long thin arrows move 
white circles on a solid straight line to black circles on a 
one-point-dotted-line curve. (A viscous liquid shows such a spatial 
gradient of the velocity. The influence of adjacent layers in a flow 
propagates along a direction perpendicular to the particle motion. Hence, the excitation 
caused by the shear viscosity is a transverse excitation.) 
Let us assume that a liquid in Fig.1 is in the Bose-Einstein Condensation (BEC) phase, and 
the many-body wave function has permutation symmetry everywhere in a capillary.  
At first sight, these displacements by long arrows seem to be a large-scale configuration change,  
 but they are reproduced by a set of slight displacements by short 
 thick arrows from the initial particles. In contrast with the longitudinal 
displacement, the transverse displacement does not change the particle 
density in the large scale, and therefore, to any given particle after 
displacement, it is always possible to find a particle being close to 
that particle in the initial configuration \cite {log}. 
  In Bose statistics, owing to permutation symmetry, one cannot 
distinguish between two types of particles after displacement, one  moved 
from the neighboring position by the short arrow, and the other moved 
from distant initial positions by the long arrow. {\it Even if  the 
displacement made by the long arrows is a large 
displacement in classical statistics, it is only a slight 
displacement  by the short arrows in Bose statistics \/}. 

Let us imagine this situation in 3N-dimensional configuration space. The 
excited state made of slight  displacements, which is a characteristic of Bose 
statistics, lies in a small distance from the ground state in configuration space.  
Since the wave function of the excited state is orthogonal to that of the 
ground state in the integral over configurations,
 the amplitude of the many-body wave function of the excited state must 
spatially oscillate around zero. Accordingly, it oscillates within a small 
distance in configuration space.  The kinetic energy of the system 
is determined by the 3N-dimensional gradient of the many-body wave 
function in configuration space, and therefore  this steep rise and fall 
of the amplitude raises the excitation energy of this wave function. 
The relaxation from such an excited state is a rapid process   
with a small $\tau $. This mechanism explains why  Bose statistics leads 
to the small coefficient of shear viscosity $\eta =G\tau $.

 When the system is at high temperature, the coherent wave function has a microscopic size.  
 If a long arrow in Fig.1 takes a particle to a position beyond the coherent wave function 
 including that particle, one cannot regard the particle after 
 displacement as an equivalent of the initial one.  The mechanism below 
$T_{\lambda}$ does not work for the large  displacement extending over two different wave 
functions.  Hence, the relaxation time $\tau $ changes to an ordinary 
long $\tau $, which is characteristic of a normal liquid \cite {ber}.

\section{shear viscosity above $T_{\lambda}$}
 Figure 4 shows $\nu (T)$ of liquid helium 4 in the vicinity of  $T_{\lambda}$ \cite {bar}. 
 One notes that  $\nu (T)$ does not abruptly drops to zero at  $T_{\lambda}$. Rather, after reaching a 
 maximum value at 3.7K, it gradually decreases with decreasing 
temperature, and finally drops to zero at the $\lambda$-point.
 $\nu (T)$ of a classical liquid ($\nu _n$ in Eq.(24)) has the general property of 
 increasing monotonically with decreasing temperature \cite {she}.
In Fig.4,  $\nu $ above $3.7K$ seems to follow this property, whereas its gradual 
fall below $3.7K$ suggests $\hat {\rho _s}(T)\ne 0$ in Eq.(24) at 
$T_{\lambda}\leq T\leq 3.7K$ \cite {spe}.

\begin{figure}
\includegraphics [scale=0.5]{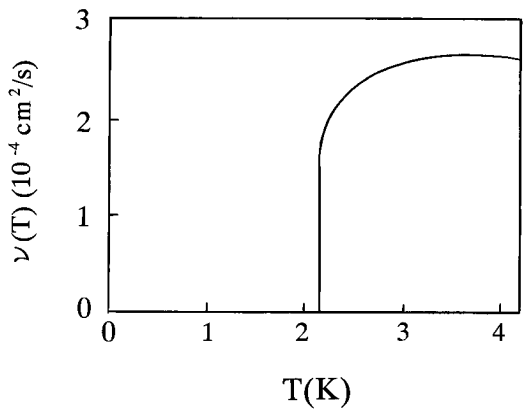}
\caption{\label{fig:epsart} 
   The temperature dependence of the kinematic shear viscosity $\nu (T)$ of liquid 
  helium 4 obtained in Ref.\cite {bar}. }
\end{figure}

In an ideal Bose gas, one knows $\hat {\rho _s}(T)=0$ at $T>T_{\lambda}$, hence
 obtains $\nu (T)=\nu _n$ at $T>T_{\lambda}$ in Eq.(24). This means 
 that, without the interaction between particles,  
BEC is the necessary condition for the superfluid flow.
 To explain the gradual fall of $\nu (T)$ just above $T_{\lambda}$, we must obtain  
$\chi^L(q,\omega)-\chi^T(q,\omega)$ under the repulsive interaction $U$. 
  The susceptibility is decomposed into the longitudinal and transverse part 
($\mu ,\nu =x,y,z$) 
\begin{equation}
	\chi_{\mu\nu}(q,\omega )=\frac{q_{\mu}q_{\nu}}{q^2¥}\chi^L(q,\omega)     
	            +\left(\delta_{\mu\nu}-\frac{q_{\mu}q_{\nu}}{q^2¥}\right)¥\chi^T(q,\omega) .
	\label{¥}
\end{equation}¥
For the later use, we define a  term proportional to $q_{\mu}q_{\nu}$ in $\chi_{\mu\nu}$ by 
 $\hat{\chi}_{\mu\nu}=(q_{\mu}q_{\nu}/q^2)[\chi^L(q,\omega)-\chi^T(q,\omega)]$  
 for considering $\hat {\rho _s}(T)=\lim_{q\to 0}[q^2/(q_{\mu}q_{\nu})] \hat {\chi} _{\mu\nu}$.
 Under the repulsive interaction $\hat{H_I}(\tau)$, $\hat{\chi}_{\mu\nu}$ is derived from  
 \begin{equation}
		<G|T_{\tau}J_{\mu}(x,\tau)J_{\nu}(0,0)|G>  
       =  \frac{\displaystyle{<0|T_{\tau}\hat{J}_{\mu}(x,\tau)\hat{J}_{\nu}(0,0)
 	              exp\left[-\int_{0}^{\beta¥}d\tau \hat{H}_I(\tau)¥\right]|0>¥}}
 	        {\displaystyle{<0|exp\left[-\int_{0}^{\beta¥}d\tau  \hat{H}_I(\tau)¥\right]|0>¥}}¥,
	\label{¥}
\end{equation}¥
where $J_{\mu}(q,\tau)=\sum_{p,n} (p+q/2)_{\mu}\Phi_p^{\dagger}\Phi_{p+q}e^{-i\omega _n\tau}¥$, 
 ($\beta =1/(k_BT), \tau =it$).
 Compared to an ordinary liquid, liquid helium 4 above $T_{\lambda}$ 
 has a  $10^{-3}$ times smaller coefficient of shear viscosity as shown 
 in Fig.4.  Although in the normal phase, 
it is already an anomalous liquid  under the strong influence of Bose statistics.
 Hence, it seems  natural to assume that, as  $T\rightarrow T_{\lambda}$ 
 in the normal phase, large but not yet macroscopic coherent 
 wave functions gradually grow as a thermal equilibrium state,
 and suppress the shear viscosity \cite {lod}.
 
  \begin{figure}
\includegraphics [scale=0.59]{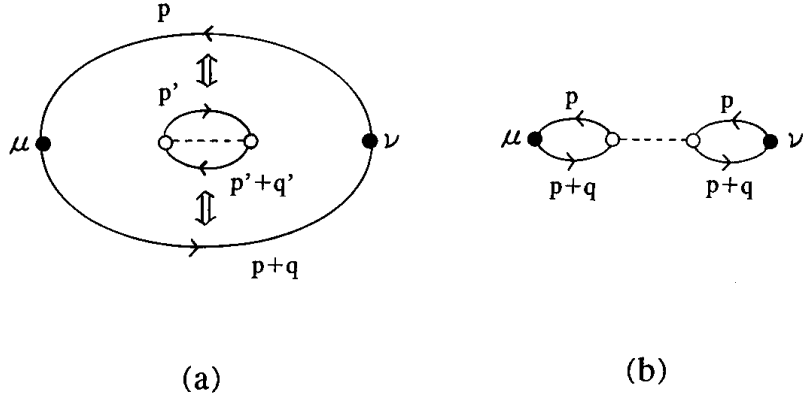}
\caption{\label{fig:epsart} Exchange of particles in (a)
         between $J_{\mu}J_{\nu}$ (a large bubble) and a bubble  
           excitation (an inner small one) yields (b). 
            }
\end{figure}

Here, we summarize the result of Ref.\cite {koh}, which formulates the 
interpretation in Sec.2B. Figure 5 illustrates  the current-current response tensor 
 $\hat{J}_{\mu}(x,\tau)\hat{J}_{\nu}(0,0)$ (a large bubble with $\mu $ 
 and $\nu $) in the liquid:  Owing to $exp(-\smallint \hat{H}_I(\tau)d\tau)$, 
 scattering of particles frequently occurs as illustrated by an 
 inner small bubble with a dotted line $U$. 
 As the  order of the perturbation increases, $\hat{J}_{\mu}(x,\tau)\hat{J}_{\nu}(0,0)$ 
in the vicinity of $T_{\lambda}$ gradually reveals a new effect due to Bose statistics:
 The large bubble $\hat{J}_{\mu}(x,\tau)\hat{J}_{\nu}(0,0)$ and the small bubble 
  in Fig.5(a) form a coherent wave function as a whole \cite {com}: 
  When one of the two particles in the large  bubble and in the small bubble have the same momentum 
($p=p'$), and when the other in both bubbles have another same momentum 
($p+q=p'+q'$), Bose statistics forces us to include in the expansion a graph made by 
exchanging these particle. This exchange yields Fig.5(b), in which  
 two bubbles with the same momenta are linked by the repulsive interaction.
 With decreasing temperature, the coherent wave function grows, 
 and such an exchange of particles occurs many times.  
Furthermore,  among possible physical processes,  processes 
including $p=0$ particles get to play a more  
dominant role than others: In Fig.5, a bubble with $p=0$ corresponds to an excitation from the rest 
particle, and a bubble with $p=-q$ corresponds to a decay into the rest 
one.  Taking only such processes and continuing these exchanges, one 
obtains $\hat{\chi}_{\mu\nu}(q,0)$ for $\hat {\rho _s}(T)$
\begin{equation}
	\hat{\chi}_{\mu\nu}(q,0)=\frac{q_{\mu}q_{\nu}}{2¥}¥
	                 \frac{1}{V¥}\frac{F_{\beta}(q)}{1-UF_{\beta}(q)¥},¥¥
	\label{¥}
\end{equation}¥
where
\begin{equation}
	 F_{\beta}(q)= \frac{(\exp(\beta[\Sigma-\mu])-1)^{-1}-(\exp(\beta[\epsilon (q)+\Sigma-\mu])-1)^{-1}} 
	                         {\epsilon (q)¥¥} ¥
	\label{¥}
\end{equation}¥
 is a positive and monotonically decreasing function of $q^2$, which approaches zero as 
$q^2\rightarrow \infty$.  An expansion of $F_{\beta}(q)$ around $q^2=0$,
 $F_{\beta}(0)-bq^2+\cdots$ has a form such as
\begin{equation}
	 F_{\beta}(q) =\frac{\beta}{4\sinh ^2 \displaystyle{\left(\frac{|\beta[\mu(T)-\Sigma]}{2¥}\right)}¥¥¥}
	                   \left[1-\frac{\beta}{2¥}\frac{1}{\tanh  \displaystyle{\left(\frac{|\beta[\mu(T)-\Sigma]|}{2¥}¥\right)}¥¥}
	                        \frac{q^2}{2m¥}¥¥¥¥  +\cdots    \right]¥¥  .
	\label{¥}
\end{equation}¥

(a) With decreasing temperature, $\mu$ gradually approaches $\Sigma$.  As $\mu 
-\Sigma\rightarrow 0$, $F_{\beta}(q)$ in Eq.(28) increases, and it 
creates a divergence in Eq.(27): This divergence first occurs at $q^2=0$,  
because $F_{\beta}(q)$ is a positive decreasing function of $q^2$.  For a small $q$, $1-UF_{\beta}(q)¥$ in 
 Eq.(27) is approximated as $[1-UF_{\beta}(0)]+Ubq^2¥$.
 As $\mu -\Sigma\rightarrow 0$, $UF_{\beta}(0)$ increases and finally reaches 1, that is,  
\begin{equation}
     U\beta=4\sinh ^2\left(\frac{\beta[\mu (T)-\Sigma(U)]}{2¥}¥\right)¥ .
	\label{¥}
\end{equation}¥
At this point, the denominator in Eq.(27) gets to begin with $q^2$, and 
$\hat{\chi}_{\mu\nu}(q,0)$ therefore has a form of 
$q_{\mu}q_{\nu}/q^2$  at $q\rightarrow 0$, the  coefficient of which is 
 $F_{\beta}(0)/(2VUb)$.  By the definition of $\hat {\rho _s}(T)=\lim _{q\to 
 0}(q^2/q_{\mu}q_{\nu}) \hat {\chi} _{\mu\nu}$, one obtains
\begin{equation}
	\hat {\rho _s}(T)=\frac{1}{V¥}\frac{m}{\sinh |\beta [\mu(T)-\Sigma] |¥}¥,
\end{equation}¥
 with the aid of Eqs.(29) and (30). Here, we call $T$ satisfying Eq.(30) 
 the onset temperature $T_{on}$ of the nonclassical behavior. 

(b) In the vicinity of $T_{\lambda}$, Eq.(30) is approximated as 
$U\beta=\beta^2[\mu (T)-\Sigma(U)]^2$ for a small $\mu-\Sigma$. This 
 condition has two solutions $\mu (T)=\Sigma(U)\pm \sqrt{Uk_BT}$.  
The repulsive Bose system is generally assumed to undergo BEC 
  as well as a free Bose gas. Hence, with decreasing temperature, 
  $\mu (T)$ of repulsive Bose system should reach $\Sigma (U)$ at a finite temperature, 
  during which course the system necessarily passes a 
 state satisfying  $\mu (T)=\Sigma (U)-\sqrt{Uk_BT}¥$. 
 Consequently, $T_{on}$ is always above $T_{\lambda}$, and  
 $\nu (T)$ deviates from $\nu _n$ just above $T_{\lambda}$ in Eq.(24).

(c) As $T$ approaches $T_{\lambda}$ from $T_{on}$,  $\hat {\rho _s}(T)$ approaches
  $(m/V)[\exp (\beta [\mu(T)-\Sigma ])-1]^{-1}$. 
 This means that at $T=T_{\lambda}$, $\hat{\rho _s}(T_{\lambda})$ agrees with the 
conventional thermodynamical superfluid density $\rho 
_s(T_{\lambda})$, and it abruptly grows to a macroscopic number. 
 
(d) At $T_{\lambda}<T<T_{on}$, Eq.(31) serves as an interpolation formula for  
$\hat {\rho _s}(T)$ \cite {pha}.

\section{Comparison to experiments}
At an early stage in the study of superfluidity, $\eta (T)$  
was measured by experiments using the capillary flow \cite 
{men}\cite {tje}\cite {zin}. But early measurements have been superseded 
by that using more accurate techniques such as a vibrating wire \cite 
{goo}\cite {wel} \cite {bru}. 
 Above the $\lambda$ point, these  different experimental methods give us 
 quantitatively similar data on $\eta (T)$. Using these accumulated data, 
 a precise curve of $\eta (T)$ and $\nu (T)$ above $T_{\lambda}$ was 
obtained by statistical analysis such as cubic spline fits \cite {bar}. 
We will compare our formula with the results in Ref \cite {bar}.

\begin{figure}
\includegraphics [scale=0.5]{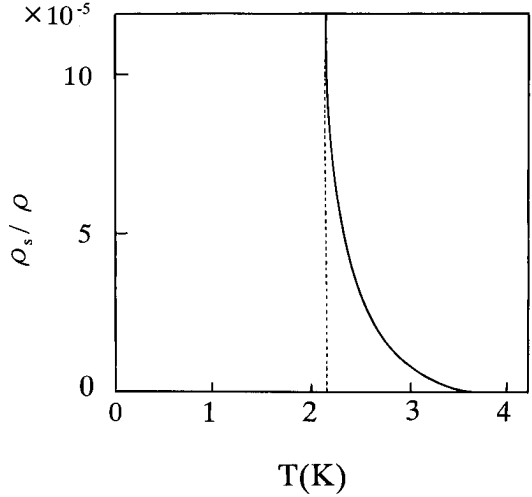}
\caption{\label{fig:epsart} 
 $\hat {\rho¥}_s(T)/\rho$ obtained by Eq.(32) using $\nu (T)$ in 
  Fig.4.  }
\end{figure}

(1) In Eq.(24), we assume $T_{on}=3.7K$ and $\nu _n=\nu (T_{on})$.  
We obtain $\hat {\rho¥}_s(T)/\rho$ by
\begin{equation}
 \frac{8}{\pi a^2\omega_s¥}¥\frac{\hat {\rho¥}_s(T)}{ \rho¥}¥ =\nu (T)^{-1}-\nu (T_{on})^{-1}.
	\label{¥}
\end{equation}¥ 
In the rotation experiment by Hess and Fairbank \cite {hes}, the 
 moment of inertia $I_z$ just above $T_{\lambda}$ is slightly smaller than the  normal 
 phase value $I_z^{cl}$ (see Sec.5.A).  Using these data,  Ref.\cite 
 {koh} roughly estimates $\hat {\rho¥}_s(T)/\rho$ as
  $\hat{\rho _s}(T_{\lambda}+0.03K)/\rho\cong 8\times 10^{-5}$, and  
 $\hat{\rho _s}(T_{\lambda}+0.28K)/\rho\cong 3\times 10^{-5}$. 
 The previous experiments in Ref.\cite {men}\cite {tje}\cite {zin} differ in the capillary radius $a$. 
 If we use a typical value $a\simeq 5\times 10^{-3}cm$, with these $\hat 
 {\rho¥}_s(T)/\rho$ in Ref \cite {koh} and $\nu (T)$ in Fig.4, we obtain a rough  
 estimate of $\omega_s$ as $\omega_s \simeq 4\times 10^{-3} rad/s$. 
 Figure 6 shows the temperature dependence of $\hat {\rho¥}_s(T)/\rho$ 
 derived by using $\nu (T)$ (Fig.4) in Eq.(32).  
 (The  absolute value is adjusted to $\hat {\rho¥}_s/\rho$ at $T_{\lambda}+0.03K$ and $T_{\lambda}+0.28K$ 
 in Ref \cite {koh}. Since the precision of these values 
 derived from currently available data is limited, the absolute value of $\hat 
 {\rho¥}_s(T)/\rho$ in Fig.6 has a statistical uncertainty.)

 (2) In Sec.3, we obtained the interpolation formula for $\hat {\rho¥}_s(T)/\rho$, Eq.(31). 
 Here, we assume $\mu -\Sigma $ changes with temperature according to the formula
 \begin{equation}
	\mu (T)-\Sigma (U)=-\left(\frac{g_{3/2}(1)}{2\sqrt{\pi}¥}¥\right)^2k_BT_{\lambda}
	          \left[\left(\frac{T}{T_{\lambda}¥}\right)^{3/2}-1\right]^2,
	\label{¥}
\end{equation}¥ 
($g_{a}(x)=\sum_{n}x^n/n^{a}¥$), where the particle interaction $U$ and 
the particle density $\rho$ of liquid helium 4 are renormalized to $T_{\lambda}=2.17K$. 
  The temperature  dependence of $\hat {\rho¥}_s(T)/\rho$ derived from 
  Eqs.(31) and (33) bears a qualitative resemblance to  
 $\hat {\rho¥}_s(T)/\rho$ in Fig.6. But, although remaining very small at $T_{\lambda}+0.2K<T<3.7K$, 
it  abruptly grows at 2.3K, and reaches a macroscopic number at $T_{\lambda}$ 
 (it resembles the shape of the letter $L$). 
   Experimentally, however, as  temperature decreases from 3.7K to 
$T_{\lambda}$, $\hat {\rho¥}_s(T)/\rho$ gradually grows as in Fig.6. This 
means that Eq.(31) is too simple to compare with the real 
system. With decreasing temperature, in addition to the particle  
 with $p=0$, other particles having small but finite momenta get to 
 contribute to the $1/q^2$ divergence of  
 $\hat{\chi}_{\mu\nu}(q,0)$ as well. (In addition to Eq.(28), a new $F_{\beta}(q)$ 
 including $p \ne 0$  also satisfies $1-UF_{\beta}(0)=0$ in Eq.(27).) 
The participation of $p \ne 0$ particles into $\hat{\rho _s}(T)$ 
is a physically natural phenomenon.  For the  repulsive Bose system, particles are likely to spread 
uniformly in coordinate space due to the repulsive force. This feature 
makes the particles with $p\ne 0 $ behave similarly with other particles, 
especially with the particle having zero momentum.
 {\it If they move at different velocities along the flow direction, the 
 particle density becomes locally high, thus  raising the interaction energy.\/}  This is a 
reason why {\it many particles with $p \ne 0$ participate in a superfluid
even above $T_{\lambda}$ \/} \cite {den}. 

 (3)  For the rigidity of superfluidity, we used two types of phenomenological 
 parameters in Sec.2. Among them, $\omega_s$ of the sharp peak in 
 $\sigma_1(\omega)$ is useful for comparing the theory with
 experiments.  The quantity which has a physically clear meaning, 
 however, is $\omega _0$ in $\lim_{q\to 0}[\chi^L(q,\omega)-\chi^T(q,\omega)]$. 
As an example, we used a step function of $\omega$ with a width $\omega _0$.  
Its Hilbert transform $\sigma _1(\omega)$ in Eq.(22) shows a 
$\delta$-function-like peak  with a width $\omega _s$ such as
$\omega _s= 10^{-3}\omega _0 \sim  10^{-2}\omega _0$ in Fig.3.  Since 
$\omega _s\cong 10^{-3}rad/s $, the critical frequency $\omega _0$ above 
which the condition of superfluidity $\lim_{q\to 0}\chi^L(q,\omega)\ne\lim_{q\to  
0}\chi^T(q,\omega)$ is violated is $10^{-1}rad/s \sim 1 rad/s$. 

(4) Equations (30) and (33) with $T_{on}=3.7K$ gives us a rough estimate of $U$ as 
$U\sim 3.4\times 10^{-16}erg$, which is  somewhat larger than the value 
 obtained from the sound velocity using the Bogoliubov formula.

 \section{Discussion}
 
 \subsection{Superfluidity in the dissipative and the non-dissipative systems}   
  On the onset mechanism of superfluidity, one can see physical 
differences between dissipative and non-dissipative flows. 
As an example of non-dissipative flows, we considered the flow in a rotating bucket in Sec.2, 
in which a quantity directly indicating the onset of superfluidity is the moment of inertia
 \begin{equation}
     I_z= I_z^{cl} \left(1-\frac{\hat {\rho _s}(T)}{\rho¥}¥\right)¥ ,¥
\end{equation}¥
where $I_z^{cl}$ is its classical value. Equation (34) and (24) have the 
following differences.

(a) In Eq.(34), $\hat {\rho _s}(T)$ appears  as a correction to the 
coefficient of the linear term $I_z^{cl}$, whereas  
 Eq.(24) has a form of infinite power series of $\nu _n$ and $\hat {\rho _s}(T)$. 
With decreasing temperature, the higher-order terms become dominant in Eq.(24). 

(b) In Eq.(34),  the change of $\chi ^L-\chi ^T$ affects $I_z$ without being enhanced,  
 and therefore the small $\hat {\rho _s}(T)$ only slightly affects 
 $I_z$ above $T_{\lambda}$ \cite {koh}.
In Eq.(24),  the change of $\chi ^L-\chi ^T$ does not directly 
affect $\nu $, but through the  dispersion integral as in Eq.(20). By this 
mechanism, a small change of the system leading to superfluidity is 
enhanced to an easily observable scale, which amplification mechanism 
does not appear in superfluidity in the non-dissipative systems.  

(c)  Equation (34) includes no phenomenological parameter, whereas Eq.(24) 
includes the  parameter $\omega _s$ representing the measure of the rigidity of superfluidity. 
This feature is characteristic of superfluidity in the dissipative systems.  The 
stability of a superfluid has been studied in the context of the critical velocity of the persistent 
current \cite {rep}.  To obtain $\omega _s$ or $\omega _0$, similar 
considerations are needed for the stability of a superfluid against 
oscillating perturbations such as the oscillating velocity $v(r) e^{i\omega t}$. 
 This is a future problem.

 Anomalously high thermal conductivity of liquid helium 4 at 
 $T<T_{\lambda}$ is another example of superfluidity in the dissipative system. 
 Heat flow $q$ is expressed as $q=-\kappa\nabla T$, where $\kappa$ is 
 the  coefficient of thermal conductivity. In the critical region above 
 $T_{\lambda}$ ($T/T_{\lambda}-1<10^{-3}$), the rapid rise of $\kappa$ is  
 observed. At $T=T_{\lambda}$, $\kappa$ jumps to an at least $10^7$ times 
 higher value than $\kappa$ just above $T_{\lambda}$.  In $T_{\lambda}<T<3.7K$,  however, 
 $\kappa$ does not show a symptom of its rise, in contrast with $\eta $ 
 which shows a symptom of its fall in the same region \cite {kel}. In 
 the case of the capillary flow, we  can assume the flow in a rotating bucket 
 as a non-dissipative counterpart, whereas  
 in the case of heat conduction, we do not know such a counterpart. 
 Theoretically, the coefficient of thermal conductivity is expressed by a 
 correlation  function which has a formally similar form to that of the shear viscosity.
 But heat conduction is a transport phenomenon of energy. 
 The scalar field (temperature field) has only a small variety of spatial distribution
 compared to the vector field (velocity field), and therefore heat 
 conduction is always a dissipative phenomenon.  
 Hence, we can not apply the formalism in this paper to the onset 
 mechanism of the anomalous thermal conductivity. 
 This formal difference between the shear viscosity  and the thermal 
 conductivity is consistent with the experimental difference between them at $T_{\lambda}<T<3.7K$.

\subsection{Comparison to  Fermi liquids}   
 In  liquid helium 3, the fall of the shear viscosity at $T_c$ is known 
as a parallel phenomenon to that of  liquid helium 4.  The formalism 
in Sec.2 is applicable to the shear viscosity of  liquid helium 3 as 
well.  For the behavior above $T_c$, however, 
there is a striking  difference between   liquid helium 3 and 4. 
The phenomenon occurring in fermions in the vicinity of $T_c$ is not a 
gradual growth of the coherent wave function, but a formation of the Cooper pairs from two fermions. 
 (This difference evidently appears in the temperature dependence of the 
 specific heat: $C(T)$ of  liquid helium 3 above $T_c$ does not show a symptom of its rise.)
 Once the Cooper pairs are formed, they are  composite bosons situated at low temperature 
and high density,  and immediately jumps to the ESP or BW state. 
Hence, the shear viscosity of liquid helium 3 shows an abrupt drop at 
$T_c$ without a gradual fall above $T_c$.    
 
In electron superconductivity, the fluctuation-enhanced 
conductivity $\sigma '$  is observed above $T_c$ (see the text by M.Tinkham 
in \cite {fer}). In  bulk superconductors, 
however, the ratio of $\sigma '$ to the normal  conductivity  $\sigma _n$ 
is about $10^{-5}$ at the critical region, and zero outside of this region.  
 Practically,  it is unlikely that thermal fluctuations create a large 
 change of $\sigma$ at temperatures outside of the critical region.  
 
 \subsection{Velocimetry technique}   
Since the discovery of superfluidity in liquid helium 4, a lot of 
measurements have been done, but there is so little direct experimental 
information about the flow patterns.  Among various visualization techniques used in ordinary 
liquids, {\it particle image velocimetry\/} (PIV),  which records 
the motion of micrometre-scale solid particles suspended in the fluid as tracer 
particles, recently becomes available in superfluid helium 4. Until 
now,  this technique has been mainly used for the study of turbulent flows \cite 
{don}\cite {van}, but it has the potential to give us more information. 
   The quantity which directly indicates the onset of superfluidity in 
 dissipative system is the change of the conductivity spectrum $\sigma (\omega)$.  In view of 
the gradual decrease of $\eta $ above $T_{\lambda}$, the sharp peak 
at $\omega =0$ and the corresponding change from $\sigma (\omega)$ to $\sigma 
_n(\omega)$ in Fig.2 must already appear at $T>T_{\lambda}$.
 To confirm this prediction, under the slowly oscillating pressure, 
 time-resolved measurement of the oscillating flow 
velocity $\mbox{\boldmath $v$}(\mbox{\boldmath $r$},t)$ 
 is necessary. Such an experiment must be performed 
in a thin capillary with an inner radius of $10^{-2}\sim 10^{-1}$ mm. 
 If  the PIV experiment is performed under such conditions and the 
 data of $\mbox{\boldmath $v$}(r=0,t)$ is Fourier-transformed, the change of $\sigma 
(\omega)$ in Fig.2 will be observed.  Since quantized vortices are absent in the 
capillary flow, tracer particles  interact only with the normal fluid part 
and trace its velocity \cite {poo}. Hence, in Fig.2(b) only the change 
from $\sigma (\omega)$ to $\sigma _n(\omega)$  will be observed.  The quantity $A$ in Eq.(8) is 
equal to the area of the shaded region $\sigma (\omega)-\sigma _n(\omega)$ in Fig.2(b). Using thus 
obtained $A$, one can determine $\omega _s$  by comparing the 
experimental data of $\eta (T)$ with Eq.(9).  Furthermore, using 
$\pi a^2A=2\hat {\rho¥}_s(T)/\rho$, one will obtain 
a new estimate of $\hat {\rho¥}_s(T)/\rho$ at $T>T_{\lambda}$.
Such an experiment may be a difficult one, but it will provide us valuable information.

\appendix

 \section{Conductivity spectrum $\sigma (\omega)$}
The Stokes equation under the oscillatory pressure gradient $\Delta Pe^{i\omega t}/L$ 
is written in the cylindrical polar coordinate as follows
 \begin{equation}
 \frac{\partial v}{\partial t¥}¥=\nu \left( \frac{\partial }{\partial r^2¥}+ \frac{\partial }{r\partial r¥}\right)v
                              + \frac{\Delta Pe^{i\omega t}}{\rho¥L}¥.
\end{equation}¥
The velocity has the following form
\begin{equation}
 v(r,t)¥=\frac{\Delta Pe^{i\omega t}}{i\omega \rho¥L}+\Delta v(r,t)¥,
\end{equation}¥
with the boundary condition of $v(a,t)=0$. $\Delta v(r,t)$ satisfies 
\begin{equation}
 \frac{\partial \Delta v(r,t)}{\partial t¥}¥=\nu \left( \frac{\partial}{\partial r^2¥}
                      + \frac{\partial }{r\partial r¥}\right) \Delta v(r,t)¥,
\end{equation}¥
and therefore $\Delta v(r,t)$ has a solution written by the Bessel function $J_0(i\lambda r)$
with $\lambda =(1+i)\sqrt {\omega /(2\nu)}$. Hence,
\begin{equation}
 v(r,t)=\frac{\Delta Pe^{i\omega t}}{i\omega \rho¥L¥}¥
                       \left(1-\frac{J_0(i\lambda r)}{J_0(i\lambda a)¥}\right)¥.
\end{equation}¥
At $r=0$, the conductivity spectrum $\sigma (\omega )$, which is given by 
 $\rho v(0,t)=\sigma (\omega )a^2 \Delta Pe^{i\omega t}/L$, has the 
 following form
\begin{equation}
 \sigma (\omega )=\frac{1}{i\omega a^2¥}\left(1 
 -\frac{1}{J_0\left[ia(1+i)\displaystyle {\sqrt {\frac{\omega}{2\nu¥}¥}}\right]¥¥}\right)¥.
\end{equation}¥
The real part of Eq.(A5) gives a curve of $\sigma (\omega)$ in Fig.2(a)  
(Re $\sigma (0)=\rho /(4\eta )$),  
and determines $f(a)\propto a^{-2}$ in Eq.(7). ($\nu $ disappears in $\int \sigma (\omega)d\omega$.)
Im $\sigma (\omega)$ gives an expression of $\sigma 
_2(\omega )$ in Eq.(10), but gives no concrete form of $\sigma _2(\omega )$ 
in Eq.(16), because it is derived from the phenomenological 
 equation with dissipation like the Stokes equation.

\section{Maxwell's relation}
Consider shear transformation of a solid and of a liquid. In a solid,  
shear stress $F_{xy}$ is proportional to a shear angle $\phi$ as 
$F_{xy}=G\phi$, where $G$ is the modulus of rigidity. The value of $G$ is 
determined by dynamical processes in which vacancies in a solid move to neighboring 
positions over the energy barriers. As  $\phi$ increases, $F_{xy}$ 
increases as follows,
\begin{equation}
 \frac{dF_{xy}}{dt}=G\frac{d\phi}{dt¥}¥¥.
	\label{¥}
\end{equation}¥
In a liquid, the motion of a fluid changes the relative positions of particles, 
reducing the shear stress $F_{xy}$ to a certain value. 
Presumably, the rate of such relaxations is proportional to the magnitude of $F_{xy}$, and one obtains
\begin{equation}
 \frac{dF_{xy}}{dt}=G\frac{d\phi}{dt¥}-\frac{F_{xy}}{\tau¥}¥¥¥,
	\label{¥}
\end{equation}¥
where $\tau $ is a relaxation time. 
In the stationary flow after relaxation, $F_{xy}$ remains constant, and 
one obtains
\begin{equation}
 G\frac{d\phi}{dt¥}=\frac{F_{xy}}{\tau¥}¥¥¥.
	\label{¥}
\end{equation}¥
 Figure.7 shows two particles 1 and 2, each of which starts at $(x,y)$ and $(x,y+\Delta y)$ 
simultaneously and moves along the $x$-direction.  Assume that there is 
velocity gradient $v_x(y)$ along $y$ direction. After $\Delta t$ has 
passed, they (1' and 2') are at a distance of $\Delta v_x\Delta t$ along the 
$x$-direction. The shear angle increases from zero to $\Delta\phi$ as a 
result, which satisfies $\Delta v_x\Delta t=\Delta y\Delta\phi$ as depicted in Fig.7. Hence, we obtain
\begin{equation}
 \frac{\partial v_x}{\partial y¥}=\frac{d\phi}{dt¥}¥.
	\label{¥}
\end{equation}¥
Substituting Eq.(B4) into Eq.(B3), and comparing it with Eq.(1), one obtains 
$\eta=G\tau$ (Maxwell's relation).

 \begin{figure}
\includegraphics [scale=0.43]{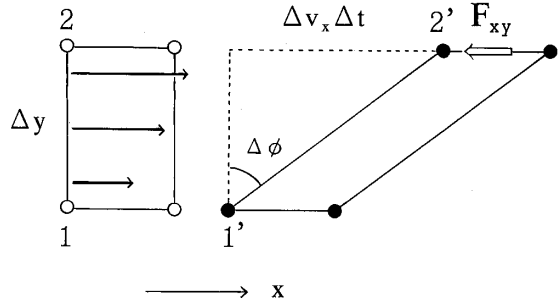}
\caption{\label{fig:epsart} In a liquid  flowing along the $x$-direction, owing to the
 velocity gradient along the $y$-direction, a small rectangular part of a liquid 
 is deformed to a parallelogram. }
\end{figure}

\newpage 

\begin{thebibliography}{99}


          \bibitem{kha} L.D.Landau and I.M. Khalatonikov, 
          Zh.Eksp.Teor.Fiz.{\bf 19\/}, 637, 709(1949),
           in {\sl Collected papers of L.D.Landau\/}, (edited by D.ter.Haar, Pergamon, London, 1965).
          
           \bibitem{kad} L.P.Kadanoff and P.C.Martin, Ann.Phys. {\bf 24\/}, 419(1963),
              P.C.Hoenberg and  P.C.Martin, Ann.Phys. {\bf 34\/}, 291(1965).
            
                      
           \bibitem{han} In the short time scale, the mechanism of the 
           shear viscosity in a liquid is similar to that in a solid, and is therefore  
          a highly inhomogeneous process at the molecular level: The modulus of rigidity $G$ 
          is determined by the motion of vacancy as in a solid. 
          (As a text, J.P.Hansen and I.R.McDonald, {\sl 
          Theory of Simple Liquids\/}, 3rd ed, (Academic Press, London, 2006)).
                       
          \bibitem{jeo} Using Eq.(2), $\eta $ of $\phi ^4$ model is 
          calculated beyond the one-loop level in the context of the relativistic heavy ion 
           collision by  S.Jeon, Phy.Rev.D.{\bf 52\/}, 3591(1995), E.Wang and U.Heinz, 
           Phys.Lett.B. {\bf 52\/}, 208(1999). These results describe the 
           shear viscosity of a strongly interacting  dense gas. 
           For a liquid, however, the anisotropic particle interaction with the 
           hard core must be introduced in calculations. 
           
           
          \bibitem{kap} P.Kapitza, Nature.{\bf 141\/}, 74(1938), 
           J.F.Allen and A.D.Misener, Nature.{\bf 141\/}, 75(1938)
        
            
           
           \bibitem{cur} As a response to $\mbox{\boldmath $P$}/L$, 
           $\mbox{\boldmath $j$}(\mbox{\boldmath $r$})$  at another 
           point is possible, but it results in the same form of $\eta$.
           
          

         
           \bibitem{can} To derive the decrease of $\eta$ from the increase of $U$ in Eq.(2), 
           delicate cancellation of higher-order terms is needed in the perturbation expansion.   
               
                        
            \bibitem{fer} Just after the advent of BCS model, an attempt was made to relate the 
             electrical conductivity (more precisely, the micro-wave absorption 
             spectrum) with the penetration depth in the Meissner effect by 
             R.A.Ferrell and R.E.Glover, Phy.Rev.{\bf 109\/}, 1398(1958), 
             and by M.Tinkham and  R.A.Ferrell, Phy.Rev.{\bf 2\/}, 331(1959). 
            (As a text,  M.TinkhamÊ {\sl Introduction to superconductivity\/}, 2nd ed, (McGraw-Hill, New York, 1996)). 
             Whereas the electrical conduction is a dissipative phenomenon, the Meissner effect is 
             a non-dissipative one, and is an analogue in superconductivity to the nonclassical 
             flow in a rotating bucket \cite {noz}. 
 
     
                
        
             
         \bibitem{sum} R.Kubo, J.Phys.Soc.Japan.{\bf 12\/}, 570(1957). 
       
         
         \bibitem{vec} In Ref \cite {fer}, the definition of the vector 
         potential $\partial A/\partial t=-E$ plays the role of Eq.(11).
          
         \bibitem{vel} $\mbox{\boldmath $v$}(\omega)$ in the right-hand side of Eq.(12) is a 
         fictitious velocity when the system would obey Eq.(11) of 
         the non-interacting system, whereas $ \mbox{\boldmath $v$}(r=0)$ included in 
         $\mbox{\boldmath $j$}(\omega)$ of the left-hand side 
         is a real velocity in the interacting system. 
           
         
            \bibitem{noz} As a review, P.Nozieres, in {\sl Quantum Fluids  \/}
            (ed by  D.E.Brewer), 1 (North Holland, Amsterdam, 1966),
          G.Baym, in {\sl Mathematical methods in 
          Solid State and Superfluid Theory \/} (ed by  R.C.Clark and 
          G.H.Derrick), 121 (Oliver and Boyd, Edingburgh, 1969)
          
  
           \bibitem{def} We use $\rho$ in the transverse response such as 
           conductivity $\sigma_{1n}=\rho /(4\eta )$ despite of $\rho$ being a longitudinal 
           susceptibility satisfying $\rho=mn=\chi^L(q,0)$, which 
           non obvious situation is explained by the replaceability of $\chi^T$ by $\chi^L$ in Eq.(19).
  
           \bibitem{ker}  E.C.Kerr, J.Chem.Phys.{\bf 26\/}, 511(1957), and 
                        E.C.Kerr and R.D.Taylor, Ann.Phys. {\bf 26\/}, 292(1964).
    
         
          \bibitem{max} J.C.Maxwell, Phil.Trans.Roy.Soc.{\bf 157\/}, 49(1867)
              in {\sl The scientific papers of J.C.Maxwell\/}, (edited by 
              W.D.Niven, Dover, New York, 2003) Vol.2, 26.
          
      
         
         \bibitem{fey1} R.P.Feynman, in {\sl Progress in Low Temp Phys\/}. 
            {\bf  1\/}, (ed C.J.Gorter), 17 (North-Holland, Amsterdam, 1955).
        
            
           \bibitem{log} For the longitudinal displacement, the large-scale 
        inhomogeneity occurs in the particle density, and therefore it 
        is not always possible to find such particles as in the transverse one. 
       
        
          \bibitem{ber}  The structural characteristic of liquids is   
           {\it irregular and only slightly different arrangements \/} of molecules. The energies $E$ of 
           these  similar structures differ only slightly to each other. 
          In the decays from one of these arrangements to others, the small $|E-E_0|$
           leads to the long $\tau $ of a normal liquid. 
          
      
       
          \bibitem{bar} C.F.Barenghi, P.J.Lucas, and R.J.Donnelly, J.Low.Temp.Phys.{\bf 44\/}, 491(1981)
        
      
       
          \bibitem{she} In a classical liquid, $\eta $ is inversely proportional to the rate of
           processes in which  vacancies in a liquid propagate from one point to 
          another over the energy barriers. With decreasing temperature, this rate 
          decreases. The liquid therefore gets to slowly respond to perturbations, and its $\eta $ increases. 
               
        
         \bibitem{spe} This hypothesis is consistent with the specific heat 
         of liquid helium 4, which shows a symptom of its rise at  $T_{\lambda}<T<2.8K$.   
               
        \bibitem{lod} London stressed that the essence of superfluidity is not the absence of viscosity, 
            but the occurrence of $rot \mbox{\boldmath $v$}=0$.   
            (F.LondonÊ {\sl Superfluid\/}, John Wiely and Sons, New York, 1954 Vol.2, 141.)  
             Although  this remark is correct, it does not rule out the 
              possibility that, behind the substantial decrease of $\eta $ 
            above $T_{\lambda}$, $rot \mbox{\boldmath $v$}=0$ is locally 
            realized by the intermediate-sized coherent wave function. Since  
            ordered motions of many particles are necessary for $rot \mbox{\boldmath $v$}=0$,  
            we cannot  attribute $rot \mbox{\boldmath $v$}=0$ to the short-lived random 
            fluctuations \cite {koh}.
             (R.Bowers and K.Mendelssohn discussed a similar view on the 
             nature of the decrease of $\eta $ above $T_{\lambda}$ \cite {men}.)
             
             
         \bibitem{koh} S.Koh, Phy.Rev.B.{\bf 74\/}, 054501(2006). 
     

            
           
            \bibitem{com} It is possible that  more complex diagrams than a bubble exchange 
             particles with the tensor $J_{\mu}J_{\nu}$.  While it is 
             difficult to estimate an infinite sum of these diagrams, it adds only a small correction. 
            
         
         \bibitem{pha}   R.A.Farrell, N.Menyhard,  H.Schmidt, F.Schwabl and P.Szepfalusy, Ann.Phys. {\bf 47\/}, 
         565(1968) considered the nonlocal superfluid density. This 
         quantity, influenced by the phase fluctuation of the order parameter,   
         only slightly deviates from the conventional  $\rho¥_s(T)$ 
         within the range of $|T/T_{\lambda}-1|<10^{-2}$. Our $\hat 
         {\rho¥}_s(T)$ at $T_{\lambda}<T<3.7K$ is a different concept from this one. 
      
      
           \bibitem{men} R.Bowers and K.Mendelssohn, Proc.Roy.Soc.Lond.A.{\bf 204\/}, 366(1950)
        
          \bibitem{tje} H.Tjerkstra, Physica.{\bf 19\/}, 217(1953)
          
          \bibitem{zin} K.N.Zinoveva, Zh.Eksp.Teor.Fiz.{\bf 34\/}, 609(1958)
                      [Sov.Phys-JETP.{\bf 7\/}, 421(1958)]
           
           \bibitem{goo} J.M.Goodwin, PhD Thesis, University of Washington (1968)
           
            \bibitem{wel} B.Welber, Phy.Rev.{\bf 119\/}, 1816(1960),  
            B.Welber and D.C.Hammer, Phys.Lett.{\bf 15\/}, 233(1965), 
            B.Welber and G.Allen, Phys.Lett.{\bf 33A\/}, 213(1970).  
            
           \bibitem{bru} L.Bruschi, G.Mazzi, M.Santini and G.Torzo, J.Low.Temp.Phys.{\bf 29\/}, 63(1977) 
          
          \bibitem{hes} G.B.Hess and W.M.Fairbank, Phy.Rev.{\bf 19\/}, 216(1967)
       
           \bibitem{den} The total $\hat{\rho _s}(T)$ is a sum of each $\hat{\rho _s}(T)$ over different momenta.
            To avoid the rise of the repulsive interaction energy, fluid particles moving along the flow direction
            participate in $\hat{\rho _s}(T)$. Hence, the density of states for this sum is not proportional
           to $4\pi p^2$.  Rather, the one-dimensional momenta along the flow direction leads to a constant 
           density of states.

        
           \bibitem{rep} J.S.Langer and J.D.Reppy,  in {\sl Progress in Low Temp Phys\/}. 
            {\bf  6\/}, (ed C.J.Gorter), ch1 (North-Holland, Amsterdam, 1970). 
               
          
       
           \bibitem{kel} L.J.Challis and J.Wilks, in   {\sl Proceedings 
           of the Symposium on Solid and Liquid $He^3$ \/}  (Ohio State 
           University, Ohio, 1957). J.F.Kerrisk and W.E.Keller,  Phy.Rev.{\bf 177\/}, 341(1969).      
      
           
              \bibitem{don} R.J.Donnelly,A.N.Karpetis, J.J.Niemela, 
              K.R.Sreenivasan, W.F.Vinen and C.M.White, J.Low.Temp.Phys. {\bf 126\/}, 327(2002),
             
             
           \bibitem{van} T.Zhang, D.Celik and S.W.Van Sciver, J.Low.Temp.Phys. {\bf 134\/}, 985(2004),
              T.Zhang and S.W.Van Sciver, J.Low.Temp.Phys. {\bf 138\/}, 865(2005)       
          
          \bibitem{poo} D.R.Poole, C.F.Barenghi, Y.A.Sergeev and W.F.Vinen, Phy.Rev.B.{\bf 71\/}, 064514(2005).   
          
           
        
      

\end{thebibliography}

\end{document}